\documentclass[aps,twocolumn,showpacs,preprintnumbers,superscriptaddress,longbibliography]{revtex4-1}

\usepackage{graphicx}  
\usepackage{subfigure}
\usepackage{multirow}

\usepackage{fancyhdr}
\usepackage{longtable}
\usepackage{parskip}
\usepackage[T1]{fontenc}
\usepackage{dcolumn}   
\usepackage{afterpage}
\usepackage{bm}        
\usepackage{amsfonts}  
\usepackage{amsmath}   
\usepackage{amssymb}   
\usepackage{physics}
\usepackage{svg}
\usepackage{hyperref}
\usepackage{mathtools}
\usepackage{float}

\AtBeginDocument{%
    \newwrite\bibnotes
    \def\bibnotesext{Notes.bib}
    \immediate\openout\bibnotes=\jobname\bibnotesext
    \immediate\write\bibnotes{@CONTROL{REVTEX41Control}}
    \immediate\write\bibnotes{@CONTROL{%
    apsrev41Control,author="08",editor="1",pages="1",title="0",year="1"}}
     \if@filesw
     \immediate\write\@auxout{\string\citation{apsrev41Control}}%
    \fi
}%

\newcommand{\pwisein}{\left\{ \begin{array}{ll}}
\newcommand{\pwiseout}{\end{array}\right.}

\usepackage{indentfirst}
\begin{document}

\title{Full counting statistics of the photocurrent through a double quantum dot embedded in a driven microwave resonator}

\author{Drilon Zenelaj}

\affiliation{\it Physics Department and NanoLund, Lund University, Box 118, 22100 Lund, Sweden}

\author{Patrick P. Potts}

\affiliation{\it Department of Physics, University of Basel, Klingelbergstrasse 82, 4056 Basel, Switzerland}

\author{Peter Samuelsson}

\affiliation{\it Physics Department and NanoLund, Lund University, Box 118, 22100 Lund, Sweden}

\date{\today}

\begin{abstract}  
Detection of single, itinerant microwave photons is an important functionality for emerging quantum technology applications as well as of fundamental interest in quantum thermodynamics experiments on heat transport. In a recent experiment [W. Khan \textit{et al.}, \href{https://doi.org/10.1038/s41467-021-25446-1}{Nat. Commun. \textbf{12}, 5130 (2021)}], it was demonstrated that a double quantum dot (DQD) coupled to a microwave resonator can act as an efficient and continuous photodetector by converting an incoming stream of photons to an electrical photocurrent. In the experiment, average photon and electron flows were analyzed. Here we theoretically investigate, in the same system, the fluctuations of the photocurrent through the DQD for a coherent microwave drive of the resonator. We consider both the zero-frequency full counting statistics as well as the finite-frequency noise (FFN) of the photocurrent. Numerical results and analytical expressions in limiting cases are complemented by a mean-field approach neglecting dot-resonator correlations, providing a compelling and physically transparent picture of the photocurrent statistics. We find that for ideal, unity efficiency detection, the fluctuations of the charge current reproduce the Poisson statistics of the incoming photons, while the statistics for non-ideal detection is sub-Poissonian. Moreover, the FFN provides information of the system parameter dependence of detector short-time properties.  Our results give novel insight into microwave photon-electron interactions in hybrid dot-resonator systems and provide guidance for further experiments on continuous detection of single microwave photons.

\end{abstract}


\maketitle 

\section{Introduction}

Photons are the elementary quanta of the electromagnetic field and have been a central concept in physics for over a century. The ability to experimentally create and detect individual photons \cite{eisaman} has found applications in areas ranging from particle physics and astronomy \cite{huber} to imaging \cite{ripoll} and spectroscopy \cite{michalet}. During the last few decades, single-photon detectors in the optical regime have attracted increasing interest due to their potential applicability in emerging quantum technologies, such as linear optics quantum computation \cite{knill}, quantum random number generation \cite{herrero} and quantum cryptography and key distribution \cite{pirandola}. In state-of-the-art photodetectors, key properties such as near-unity detection efficiency \cite{pernice}, low dark-count rates \cite{hadfield}, photon-number resolved detection \cite{kardynal} and high operation speed \cite{ackert} have been demonstrated.  

In the microwave regime, photons have an energy which is four to five orders of magnitude smaller than in the optical regime. This puts very different requirements for experimental investigations and applications based on microwave photons \cite{casariego}. During the last decade, a large number of theoretical proposals for single-microwave-photon detectors have been presented, see e.g. \cite{romero,chen,fan,sankar,kyriienko,gu}. In recent years, single-microwave-photon detection has also been demonstrated experimentally \cite{inomata,narla,opremcak,besse,kono,huard,essig}. Functionalities such as close-to-unit efficiency photodetectors \cite{narla,opremcak,huard}, quantum non-demolition measurements \cite{besse,kono} of itinerant photons and few-photon number resolution \cite{huard,essig} have been realized in experiments based on superconducting qubits. A common property of the superconducting qubit detectors \cite{narla,opremcak,huard,besse,kono} is that they require synchronization with the photon signal and typically involve advanced qubit pulsing and read-out schemes. 

Very recently, taking an altogether different approach, continuous microwave photodetection was demonstrated in a semiconductor double quantum dot (DQD) resonantly coupled to a driven superconducting resonator \cite{khan}. The operation principle is a direct microwave analog of the photocarrier generation in conventional optical photo diodes \cite{pearsall}, since incident microwave photons generate an electron current through the DQD by inducing tunneling of electrons from the ground to the excited DQD state.  The quantum efficiency of the photon-to-electron conversion in the experiment reached 6$\%$ \cite{khan}, several orders of magnitude higher than previous experiments based on similar approaches \cite{gustavsson,liu}. 

The prospect of experimentally reaching even higher values, approaching the theoretically predicted unity efficiency \cite{vavilov}, motivates further theoretical investigations of the photodetector system. On the fundamental side, the efficient conversion of a stream of photons, bosonic particles, into a current of electrons, fermionic particles, raises interesting questions on the relation of the statistical properties of the two flows. From a more applied perspective, the short-time properties of the photodetector provide information on dead-times and detection speed, important figures-of-merit for time-resolved, single-photon detection. 

In this work, we address these questions theoretically by analyzing both the statistical distribution, or full counting statistics (FCS) \cite{belzig,kiesslich,nazarov,xu,schaller}, of the number of photoelectrons transported through the DQD during a measurement as well as the finite-frequency noise (FFN) of the photocurrent. We provide a numerical analysis, complemented by analytical results in the low and large drive limits, based on the solution of the appropriate generalized quantum master equation. Moreover, we present a mean-field formalism for both the FCS and the FFN, considerably simplifying the analysis, and discuss its limits of applicability.  As key results, we show that at unit quantum efficiency, the photoelectrons inherit the Poissonian statistics of the incoming photons. For lower efficiencies, the electron statistics is typically sub-Poissonian. We moreover, from the FFN, identify the system properties governing the detector dead time. Our results will arguably stimulate further experiments on hybrid semiconductor-resonator photodetectors. It will moreover provide the framework for further theoretical investigations into photodetection with different microwave sources, e.g. non-classical ones, as well as functionalities such as single- and photon-number resolved detection.  

This paper is organized as follows. In Sec. \ref{sandm}, we present our theoretical model to describe the DQD-resonator system and show how it can be used as an efficient photodetector. We further review the conditions that need to be met in order to reach near unit photon-to-electron conversion efficiency. In Sec. \ref{fcs}, we analyze the FCS and the FFN in the limiting cases. In Sec. \ref{mfp}, we present the results for the FCS and the FFN using a mean-field approach and compare them against the solutions of Sec. \ref{fcs} and numerical calculations. We conclude and give a brief outlook in Sec. \ref{conc}.

\section{System \& Model}\label{sandm}
\subsection{Hamiltonian for the DQD-resonator system}
\begin{figure}[htbp]
    \centering
    \includegraphics[width=0.99\columnwidth]{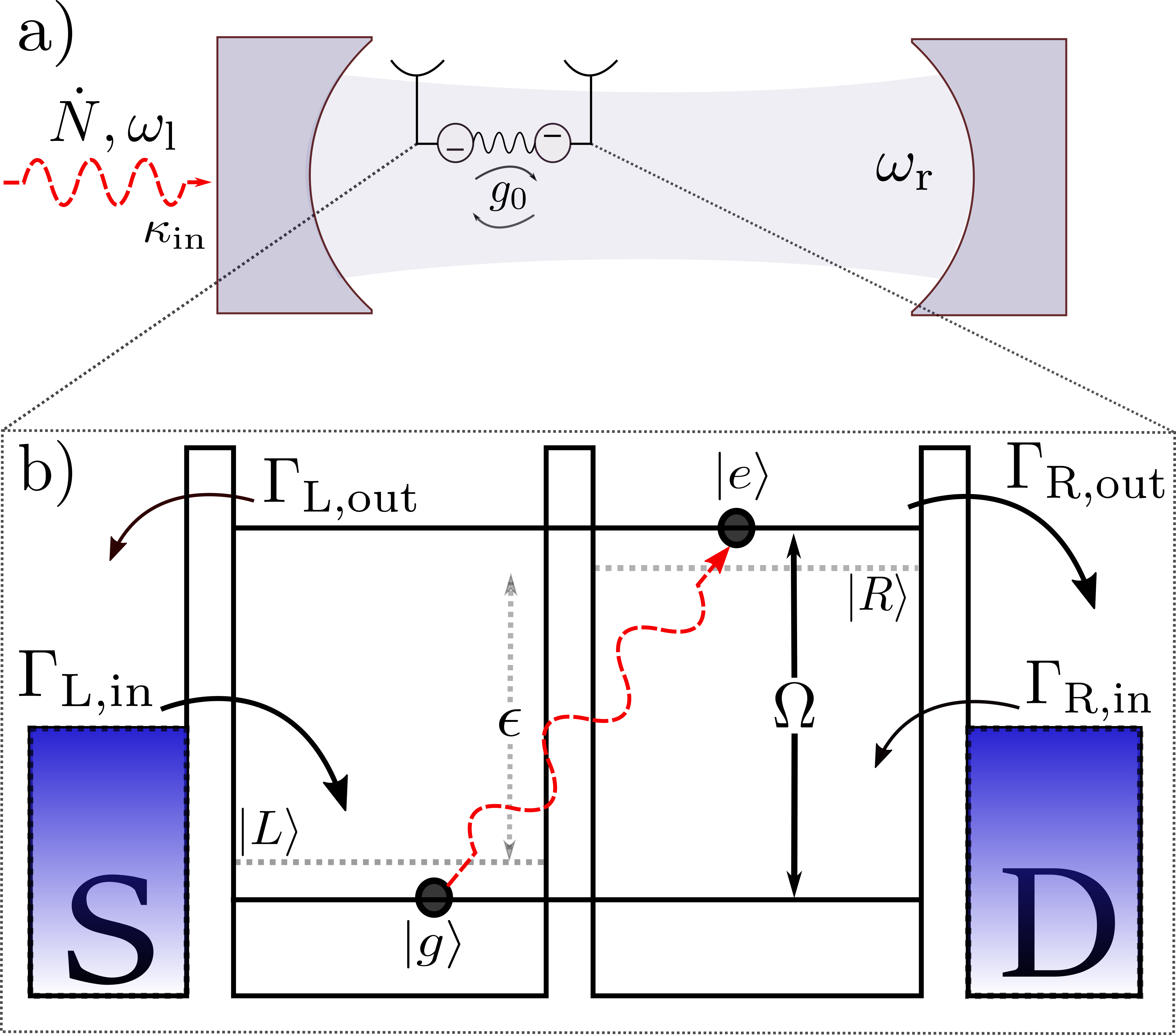}
    \caption{a) A coherent microwave drive with frequency $\omega_{\text{l}}$ and incoming photon rate $\dot{N}$ feeds in photons to a superconducting transmission line resonator with resonance frequency $\omega_{\text{r}}$. The coupling strength between drive and resonator is given by $\kappa_{\text{in}}$. The resonator is capacitively coupled to a double quantum dot (DQD) with coupling strength $g_0$. b) The DQD is tunnel coupled to a source (S) and drain (D) lead with zero applied potential bias and the levels of the DQD are tuned such that the chemical potentials of the leads lie in between the DQD ground $\ket{g}$ and excited $\ket{e}$ state, which are depicted by full black lines. The energy splitting between $\ket{g}$ and $\ket{e}$ is given by $\Omega$. The charge states of the left $\ket{L}$ and right $\ket{R}$ dot are depicted by grey dashed lines and the detuning between those charge states is given by $\epsilon$. Electrons from the leads can tunnel into $\ket{g}$ from the left and right with respective rates $\Gamma_{\text{L,in}}$ and $\Gamma_{\text{R,in}}$. If the DQD and the resonator are in resonance, transitions from $\ket{g}\to\ket{e}$ are enhanced. Electrons tunnel out from $\ket{e}$ into the leads to the left and right with respective rates $\Gamma_{\text{L,out}}$ and $\Gamma_{\text{R,out}}$.}
    \label{fig:setup}
\end{figure}

We consider a DQD which is capacitively coupled to a driven superconducting transmission line resonator \cite{cottet}, a system investigated both theoretically \cite{childress,berg,lambert,pulido,xu,bergenfeldt} and experimentally \cite{frey,petersson,viennot,stehlik,mi,stockk,bruhat} in recent years, see Fig. \ref{fig:setup}. The total system is described by the Hamiltonian  
\begin{equation}\label{eq:ham}
H=H_{\text{DQD}}+H_{\text{r}}+H_{\text{JC}}+H_{\text{d}},
\end{equation}
where $H_{\text{DQD}}$ describes the DQD, $H_{\text{r}}$ the resonator, $H_{\text{JC}}$ the coupling between the DQD and the resonator and $H_{\text{d}}$ accounts for the applied drive.

The DQD is operated in the Coulomb blockade regime, where only one excess electron is energetically allowed to reside on the DQD. The Hilbert space of the DQD is thus spanned by the states with one excess electron on the left dot $\ket{L}$, one on the right dot $\ket{R}$ and no excess electron $\ket{0}$. In this basis, the Hamiltonian of the DQD is given by (setting $\hbar = 1$)
\begin{equation}\label{hdqd}
    H_{\text{DQD}}=\frac{\epsilon}{2}(\ketbra{R}{R}-\ketbra{L}{L})+t_{\text{c}}(\ketbra{L}{R}+\ketbra{R}{L}),
\end{equation}
where $\epsilon$ is the detuning between the two charge states $\ket{L}$ and $\ket{R}$ and $t_{\text{c}}$ is the tunnel coupling between them. For the further analysis it is more convenient to work in the eigenstate basis of the DQD spanned by the ground state $\ket{g}$ and excited state $\ket{e}$ which are defined by
\begin{equation}\label{trans}
\begin{pmatrix} 
\ket{g} \\
\ket{e} 
\end{pmatrix} =
\begin{pmatrix}
\cos(\theta/2) && -\sin(\theta/2) \\
\sin(\theta/2) && \cos(\theta/2) 
\end{pmatrix}
\begin{pmatrix}
\ket{L} \\
\ket{R}
\end{pmatrix},
\end{equation}
where the mixing angle $\theta$ is given by $\cos(\theta)=-\epsilon/\Omega$ and $\Omega = \sqrt{4t_c^2+\epsilon^2}$ gives the energy splitting between the DQD eigenstates. In this basis, the Hamiltonian of the DQD can be written as
\begin{equation}\label{hdqd2}
    H_{\text{DQD}}=\frac{\Omega}{2} \sigma_z,
\end{equation}
where $\sigma_z=\ketbra{e}{e}-\ketbra{g}{g}$.

The Hamiltonian for the microwave resonator is given by 
\begin{equation}\label{hr}
    H_{\text{r}} = \omega_{\text{r}} a^{\dagger}a,
\end{equation}
where $\omega_{\text{r}}$ is the the resonator's characteristic frequency and the operator $a^{\dagger}$ ($a$) describes the creation (annihilation) of a microwave photon in the resonator. 

The interaction between the DQD and the resonator is described by a standard Jaynes-Cummings interaction Hamiltonian \cite{jaynes}
\begin{equation}\label{hjc}
    H_{\text{JC}}=g\left(a^{\dagger}\sigma_-+a\sigma_+\right),
\end{equation}
where $g=g_0\sin( \theta)$ gives the interaction strength between the DQD and the microwave photons inside the resonator with $g_0$ being the bare coupling strength and $\sigma_+=\ketbra{e}{g}=\sigma_-^{\dagger}$ is the DQD raising operator. 

The drive is a coherent source of monochromatic microwave radiation feeding in photons into the resonator and is described by \cite{khan, clerk}
\begin{equation}\label{hd}
    H_{\text{d}}= \sqrt{\kappa_{\text{in}} \dot{N}} \left(a^{\dagger}e^{i\omega_{\text{l}} t } + a e^{-i\omega_{\text{l}} t} \right),
\end{equation}
where $\kappa_{\text{in}}$ gives the coupling strength between the input port and the resonator, $\dot{N}$ is the rate of impinging photons and $\omega_{\text{l}}$ is the frequency of the microwave drive. 

Moving to the rotating frame of the incoming microwave radiation gives us \cite{jin, xu2}
\begin{equation}\label{hst}
\begin{split}
    \tilde{H} &=   \Delta_{\text{d}}\frac{\sigma_z}{2} + \Delta_{\text{r}} a^{\dagger}a + g\left(a^{\dagger} \sigma_- + a \sigma_+ \right) \\& 
    + \sqrt{\kappa_{\text{in}} \dot{N}} \left( a^{\dagger} + a \right),
\end{split}
\end{equation}
where we have used
\begin{equation}\label{eq:rot}
    \tilde{H}= U^{\dagger}H U + i(\partial_tU) U^{\dagger},
\end{equation}
with $U=\exp\left[i\omega_{\text{l}}t(a^{\dagger}a+\sigma_z/2)\right]$ and $\Delta_{\text{d}} = \Omega - \omega_{\text{l}}$ ($\Delta_{\text{r}}=\omega_{\text{r}}-\omega_{\text{l}}$) gives the detuning between the DQD (resonator) and the drive. 

The DQD is also tunnel coupled to two fermionic leads called source and drain, respectively. There is no applied bias between source and drain and the chemical potential as well as the energy of the empty state $\ket{0}$ are set to zero without loss of generality. The temperature of the leads is negligible compared to the DQD level spacing $k_BT \ll \Omega$, such that electrons can only enter into the ground state with rate
\begin{equation}\label{gg0}
\Gamma_{g0}=\Gamma_{\text{L,in}}+\Gamma_{\text{R,in}},
\end{equation}
where
\begin{equation}\label{gli}
\begin{split}
\Gamma_{\text{L,in}}&=\Gamma_{\text{L}} \cos^2(\theta/2), \\[0.2em]  \Gamma_{\text{R,in}}&=\Gamma_{\text{R}} \sin^2(\theta/2). 
\end{split}
\end{equation}
Here, $\Gamma_{\text{L}}$ ($\Gamma_{\text{R}}$) is the rate for electron tunneling events into or out of the left (right) dot.

Electrons in the excited state can tunnel back into the leads with rate
\begin{equation}\label{g0e}
\Gamma_{0e}=\Gamma_{\text{L,out}}+\Gamma_{\text{R,out}},
\end{equation}
where
\begin{equation} \label{glo}
\begin{split}
\Gamma_{\text{L,out}}&=\Gamma_{\text{L}} \sin^2(\theta/2),\\[0.2em] \Gamma_{\text{R,out}}&=\Gamma_{\text{R}} \cos^2(\theta/2).
\end{split}
\end{equation}

The DQD level spacing $\Omega$ is tuned to be in resonance with the resonator's characteristic frequency $\omega_{\text{r}}$ (meaning that we can set $\Delta_{\text{d}}=\Delta_{\text{r}}\equiv \Delta$). In that case, transitions from the ground to the excited state are enhanced.

\subsection{Master equation for the DQD-resonator system}
The dynamics of the DQD-resonator is modeled by the following Lindblad master equation \cite{lindblad, breuer}
\begin{equation}\label{lme}
\begin{split}
\partial_t \rho(t)&=-i[\tilde{H},\rho(t)] + \Gamma_{g0} \mathcal{D}[s_g^{\dagger}]\rho(t) + \Gamma_{0e} \mathcal{D}[s_e]\rho(t) \\[0.25em]&+ \frac{\gamma_{\phi}}{2}\mathcal{D}[\sigma_z]\rho(t) 
+\gamma_- \mathcal{D}[\sigma_-]\rho(t)+\kappa \mathcal{D}[a]\rho(t), 
\end{split}
\end{equation}
with the Lindblad superoperator $\mathcal{D}[x]\rho(t)= x\rho(t) x^{\dagger} - \frac{1}{2}\{x^{\dagger}x,\rho(t) \}$, where $\rho(t)$ is the density matrix of the DQD-resonator system. 

The first term in Eq. \eqref{lme} describes the unitary time evolution of the system. The second term describes tunneling of electrons from the leads into the ground state of the DQD, where $s_g^{\dagger}=\ketbra{g}{0}$.
The third term describes tunneling of electrons out of the excited state of the DQD into the leads, where $s_e=\ketbra{0}{e}$.
The fourth term describes dephasing of the state of the DQD, which is typically caused by voltage fluctuations in the electromagnetic environment \cite{buttiker}, and is quantified by the rate $\gamma_{\phi}$. The fifth term describes relaxation of an electron from the excited into the ground state and is typically due to coupling of electrons to phonons in the solid state environment \cite{grodecka, gawarecki}. Relaxation is quantified by the rate $\gamma_-$. Thermal excitations can be neglected because $k_BT \ll \Omega$. The resonator is subject to losses due to coupling to the input port of the resonator and also internal losses due to coupling to the substrate, leading to a total loss rate given by $\kappa$.

\subsection{Ideal photodetection}
\begin{figure}[htbp]
    \centering
    \includegraphics[width=0.99\columnwidth]{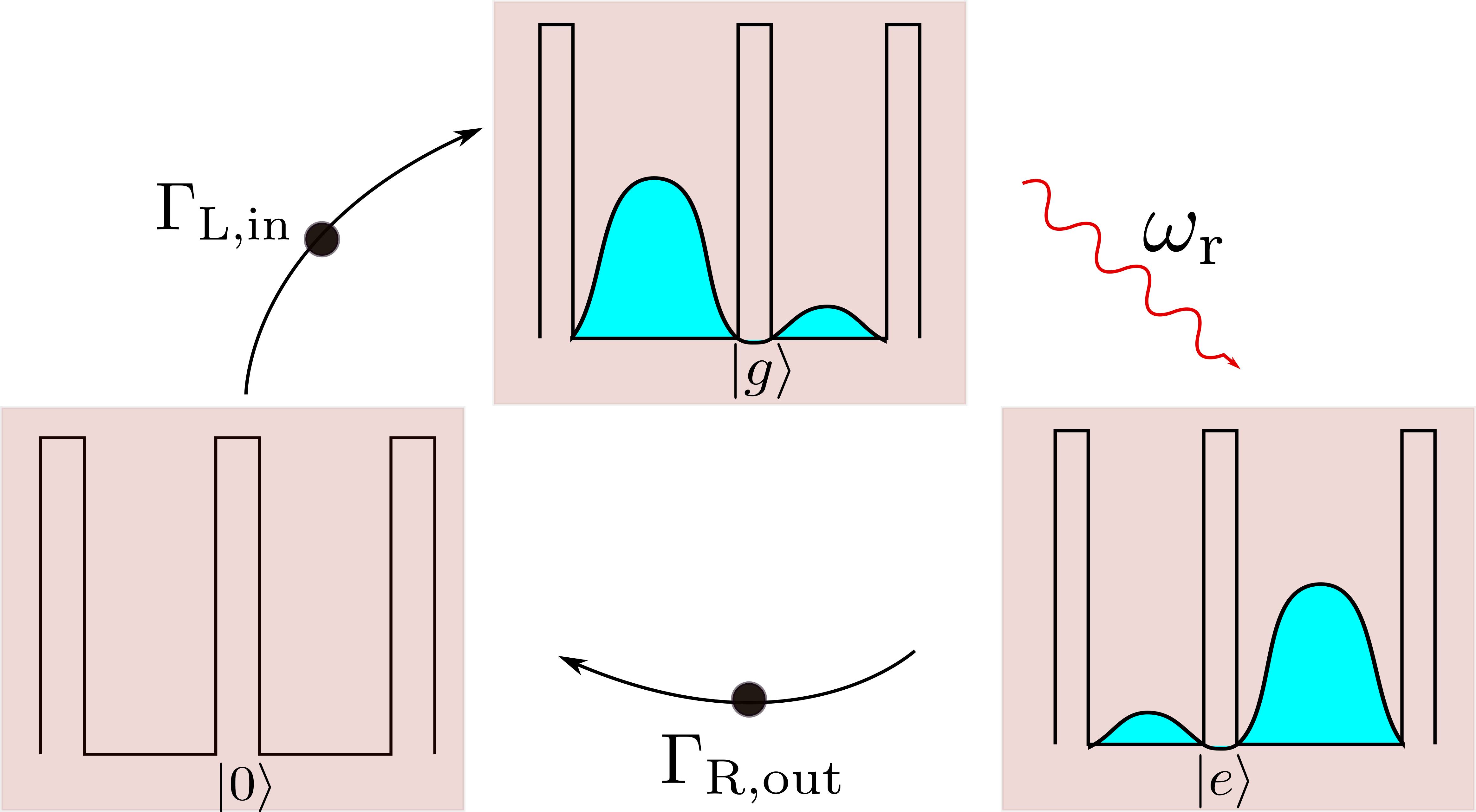}
    \caption{Schematic of an ideal photodetection cycle using a DQD coupled to a microwave resonator. The DQD starts in its empty state $\ket{0}$ where no excess electron resides on it. An electron from the source can tunnel into the ground state $\ket{g}$ with rate $\Gamma_{\text{L,in}}$. The DQD is in resonance with the resonator's characteristic frequency $\omega_{\text{r}}$, such that the electron in the ground state moves to the excited state $\ket{e}$ by absorbing a microwave photon. It will then tunnel out to the drain with rate $\Gamma_{\text{R,out}}$, which brings the DQD back to its empty state and closes the cycle.}
    \label{fig:ipd}
\end{figure}

The operation of the DQD-resonator as an ideal photodetector was described in Ref. \cite{vavilov} and is briefly discussed here for completeness. The photodetection scheme is illustrated in Fig. \ref{fig:ipd}. In the ideal case, each photon arriving at the resonator is absorbed by the DQD, giving rise to a photocurrent from source to drain. 

In practice, the photodetector will not be ideal. A relevant figure of merit to quantify how close the DQD-resonator system is to the ideal photo-detector is the photon-to-electron conversion efficiency, which is given by 
\begin{equation}\label{eff}
    \eta = \frac{I/e}{\dot{N}},
\end{equation}
where $I$ is the generated photocurrent through the DQD.

In Ref. \cite{vavilov}, certain conditions for the system have been derived in order to reach $\eta \approx 1$. First, the total photonic loss rate needs to be dominated by the coupling to the input port of the resonator, i.e. $\kappa=\kappa_{\text{in}}$. Second, the frequency of the drive should match the frequency of the resonator, such that $\Delta = 0$. Third, the tunneling rates to the left and right dot need to occur on a faster time scale than the decoherence rates, i.e. $\Gamma_{\text{L,R}} \gg \gamma_-,\gamma_{\phi}$. Fourth, the tunnel coupling strength between the dots needs to be small compared to the resonator frequency, i.e. $t_{\text{c}}\ll \omega_{\text{r}}$. Fifth, the rate at which photons get absorbed by the DQD, which is given by $4g^2/(\Gamma_{0e}+\gamma_-+2\gamma_{\phi})$, should match the photonic loss rate $\kappa$. Lastly, the generated photocurrent needs to be linearly proportional to the incoming photon flux, i.e.  $I \propto \dot{N}$. This last point reflects that the probability to have every single photon absorbed by the DQD is maximized in the limit of weak microwave drive.

\section{Full counting statistics}\label{fcs}
Due to the quantum statistical nature of the photon absorption and electron tunneling processes, the photocurrent will be subject to fluctuations in time. The statistical properties are analyzed within the framework of the FCS, which allows us to infer the probability distribution $p(n,t)$ of $n$ electrons having passed to the drain in a time span $t$. From $p(n,t)$ all the cumulants of the distribution can be obtained. 

We calculate the FCS using a master equation approach developed in Ref. \cite{bagrets}. Introducing the counting field $\chi$, the variable conjugate to $n$, we can write a master equation like the one in Eq. \eqref{lme} using the $\chi -$dependent density matrix $\rho(\chi,t)$ as
\begin{equation}\label{mec}
\begin{split}
     \partial_t \rho &= -i[\tilde{H},\rho] + \Gamma_{\text{L,in}} \mathcal{D}[s_g^{\dagger}]\rho+ \Gamma_{\text{L,out}} \mathcal{D}[s_e]\rho\\
     &+\Gamma_{\text{R,in}} (e^{-i\chi}s_g^{\dagger}\rho s_g - \tfrac{1}{2}\{s_gs_g^{\dagger},\rho\})\\&+  \Gamma_{\text{R,out}}(e^{i\chi}s_e\rho s_e^{\dagger}-\tfrac{1}{2}\{s_e^{\dagger}s_e,\rho\})\\
     &+ \frac{\gamma_{\phi}}{2}\mathcal{D}[\sigma_z]\rho +\gamma_- \mathcal{D}[\sigma_-]\rho+\kappa \mathcal{D}[a]\rho \\
     &= L(\chi)\rho,
\end{split}
\end{equation}
where $L(\chi)=L_0+e^{i\chi}J_{\text{R}}+e^{-i\chi}J_{\text{L}}$ is the $\chi-$dependent Liouvillian. The term $L_0$ contains all terms in the Liouvillian which leave the number of electrons in the drain unchanged and $J_{\text{R(L)}}$ contains the terms which increase (decrease) the number of electrons by one. For a system described by $\rho(t_0)$ at time $t_0$, the general solution to Eq. \eqref{mec} is 
\begin{equation}\label{sol}
    \rho(\chi,t)=e^{L(\chi)(t-t_0)}\rho(t_0).
\end{equation}
Moving forward, we will set $t_0=0$ and assume that the system at that time is in its steady state, i.e. $\rho(t_0=0) \equiv \rho_{\text{st}}$. 

Instead of calculating the probability distribution $p(n,t)$ directly, it is more convenient to look at the cumulant-generating function (CGF) $C(\chi,t)$ of the distribution. It can readily be obtained from the $\chi-$dependent density matrix as
\begin{equation}\label{cmf}
    e^{C(\chi,t)}=\text{Tr}\{\rho(\chi,t)\}=\text{Tr}\{e^{L(\chi)t}\rho_{\text{st}}\},
\end{equation}
The cumulants can then be found by differentiation with respect to the counting field
\begin{equation}\label{cum}
    \langle\!\langle n^k \rangle \!\rangle= \frac{\partial_k}{\partial(i\chi)^k}C(\chi,t)|_{\chi=0},
\end{equation}
where $\langle\!\langle n^k \rangle \!\rangle$ gives the $k-$th cumulant of the distribution. The CGF is related to $p(n,t)$ by inverse Fourier transformation
\begin{equation}\label{prob}
    p(n,t)=\int_{-\pi}^{\pi}\frac{d\chi}{2\pi}e^{C(\chi,t)-in\chi}.
\end{equation}
Below we will employ this recipe to compute the cumulants of the photocurrent through the DQD.
\subsection{Zero-frequency FCS}
We will first investigate the FCS in the long-time, or zero-frequency, limit. In this limit, the CGF is \cite{schaller}
\begin{equation}\label{cmf3}
    C(\chi,t)=\lambda_0(\chi)t,
\end{equation}
where $\lambda_0(\chi)$ is the eigenvalue of $L(\chi)$ with the largest real part. Finding the cumulant generating function thus becomes an eigenvalue problem in an infinite-dimensional Hilbert space. A direct analytical analysis of $\lambda_0(\chi)$, providing a clear physical picture of the photocurrent statistics, will therefore only be possible in certain limiting cases. 

\subsubsection{Low-drive limit}
Of particular interest is the regime of low-microwave drive, pointed out above as a prerequisite for ideal photodetection. In this regime, the photocurrent statistics can be obtained by perturbation theory \cite{gomez}, treating the drive Hamiltonian $H_{\text{d}}$ as a small perturbation to the DQD-resonator system. The parameter in which we make the perturbative expansion is the effective drive amplitude [cf. Eq. \eqref{hd}], which we abbreviate by $f \equiv \sqrt{\kappa_{\text{in}}\dot{N}}$. 

To make progress, we note that we can write the $\chi-$dependent density matrix in the long-time limit as $\lim_{t\to\infty}\rho(\chi,t)=\exp[\lambda_0(\chi)t]\mu_0(\chi)$ (see Appendix \ref{AA} for details). This allows us to write the master equation \eqref{mec} as 
\begin{equation}\label{link}
    \lambda_0(\chi)\tilde{\rho}(\chi)=L(\chi)\tilde{\rho}(\chi),
\end{equation}
where $\tilde{\rho}(\chi)\equiv\lim_{t\to\infty} [\rho(\chi,t)/\text{Tr}\{\rho(\chi, t)\}]$. 
Expanding Eq. \eqref{link} up to lowest order in $f$ (which is the second one) using  $\lambda_0(\chi)=\lambda_0^{(0)}(\chi)+\lambda^{(1)}_0(\chi)+\lambda^{(2)}_0(\chi)$, $\tilde{\rho}(\chi)=\tilde{\rho}^{(0)}(\chi)+\tilde{\rho}^{(1)}(\chi)+\tilde{\rho}^{(2)}(\chi)$ and $L(\chi)=L^{(0)}(\chi)+L^{(1)}(\chi)$ gives the following equations
\begin{align}
    &\lambda_0^{(0)}\tilde{\rho}^{(0)} = L^{(0)}\tilde{\rho}^{(0)}, \label{pertz}\\
    &\lambda_0^{(0)} \tilde{\rho}^{(1)}+\lambda_0^{(1)} \tilde{\rho}^{(0)} =  L^{(1)}\tilde{\rho}^{(0)} + L^{(0)}\tilde{\rho}^{(1)}, \label{pertf}\\
    &\lambda_0^{(2)} \tilde{\rho}^{(0)}+\lambda_0^{(1)} \tilde{\rho}^{(1)}+ \lambda_0^{(0)} \tilde{\rho}^{(2)}=  L^{(1)}\tilde{\rho}^{(1)} + L^{(0)}\tilde{\rho}^{(2)}.\label{perts}
\end{align}
For the expansion of $\tilde{\rho}(\chi)$ we make an ansatz where we keep terms with up to one excitation, either in the resonator or the DQD. Details on the explicit form of the normalized, $\chi-$dependent density matrix to different orders in $f$ can be found in Appendix \ref{AB}. 

Solving Eqs. \eqref{pertz}-\eqref{perts} gives $\lambda_0^{(0)}(\chi)=\lambda_0^{(1)}(\chi)=0$ and
\begin{equation}\label{l0perf}
    \lambda_0(\chi)= \Gamma_{\text{LR}}(e^{i\chi}-1)+\Gamma_{\text{RL}}(e^{-i\chi}-1),
\end{equation}
with
\begin{widetext}
\begin{equation}\label{l0perf1}
    \Gamma_{\alpha \beta}=\frac{16f^2g^2(\tilde{\Gamma}+\kappa)(4g^2+\tilde{\Gamma}\kappa)\Gamma_{\alpha,\text{in}}\Gamma_{\beta,\text{out}}}{\Gamma_{g0}[4g^2(\Gamma_{0e}+\gamma_-+\kappa)+\kappa(\Gamma_{0e}+\gamma_-)(\tilde{\Gamma}+\kappa)][(\tilde{\Gamma}^2+4\Delta^2)(\kappa^2+4\Delta^2)+8g^2(\tilde{\Gamma}\kappa-4\Delta^2)+16g^4]},
\end{equation}
\end{widetext}
where $\tilde{\Gamma}\equiv\Gamma_{0e}+\gamma_-+2\gamma_{\phi}$ is the effective dephasing rate and $\alpha,\beta\in\{\text{L, R}\}$.
Equation \eqref{l0perf} describes a bi-directional Poisson process. The corresponding low-drive probability distribution is given by [cf. Eq. \eqref{prob}]
\begin{equation}\label{eq:bidir}
    p(n,t)=\left(\sqrt{\frac{\Gamma_{\text{LR}}}{\Gamma_{\text{RL}}}}\right)^n I_n(2\sqrt{\Gamma_{\text{LR}}\Gamma_{\text{RL}}}t) e^{-(\Gamma_{\text{LR}}+\Gamma_{\text{RL}})t},
\end{equation}
where $I_n(x)$ is the modified Bessel function of the first kind \cite{temme}.

Applying the conditions for ideal photodetection introduced in section \ref{sandm} to Eq. \eqref{l0perf1}, the rate $\Gamma_{\text{RL}}$ will go to zero and $\Gamma_{\text{LR}}$ will reduce to $\dot{N}$. Thus, Eq. \eqref{l0perf} reduces to
\begin{equation}\label{cgff}
    \lambda_0(\chi) = \dot{N} (e^{i\chi}-1),
\end{equation}
with a corresponding uni-directional Poissonian probability distribution
\begin{equation}\label{probf}
    p(n,t)= e^{-\dot{N}t}\frac{(\dot{N}t)^n}{n!}.
\end{equation}
This shows that the statistics of the electrons arriving to the drain in the limit of ideal photodetection is equivalent to the statistics of photons from a coherent light source \cite{walls}. The ideal photodetector does therefore not only detect each single photon but also preserves their long-time statistics.
\subsubsection{Large-drive limit}
We also find analytical results in the regime where the microwave drive is large. In this regime,
the resonator photon state has negligible quantum fluctuations and is independent on the presence of the DQD. We can then replace the photonic annihilation operator $a$ with a c-number given by 
$-2if/(\kappa + 2i\Delta)$. This replacement reduces the Liouvillian to a $5\times5$ matrix in the basis of the DQD $\rho_{\text{DQD}}=(\rho_0,\rho_g,\rho_e,\rho_{eg},\rho_{ge})^T$, where $\rho_i=\expval{\rho}{i}$ and $\rho_{ij}=\mel{i}{\rho}{j}$ (see Appendix for \ref{AAA} details). We derive an analytical expression for $\lambda_0(\chi)$ by finding the eigenvalue of the corresponding $\chi$-dependent Liouvillian in the limit $f\to \infty$  
\begin{widetext}
\begin{equation}\label{lf}
    \lambda_0(\chi)=\frac{1}{4}\left\{-\Gamma_{0e}-2\Gamma_{g0}+\sqrt{8[(e^{i\chi}-1)\Gamma_{\text{L,in}}\Gamma_{\text{R,out}}+(e^{-i\chi}-1)\Gamma_{\text{R,in}}\Gamma_{\text{L,out}}]+(\Gamma_{0e}+2\Gamma_{g0})^2}\right\}.
\end{equation}
\end{widetext}
We see that in the limit of large drive, only the in- and out-tunneling rates of the electrons are relevant for the statistics of the photocurrent. The resonator essentially works as a mediator for transitions between the ground and excited state of the DQD, which are so fast that the corresponding rates disappear from the FCS. 
\subsection{Finite-frequency noise}
The zero-frequency cumulants give us insight into many of the statistical properties of the electron transport through our system. However, a full picture of the relevant correlations and time scales of said properties requires an analysis over the full frequency range \cite{ubbelohde}. We will thus turn to the FFN, which can be calculated by employing MacDonald's formula \cite{macdonald}
\begin{equation}\label{mcd}
    S(\omega) = \omega \int_0^{\infty} dt \sin(\omega t) \frac{d}{dt} \langle\!\langle n^2 \rangle\!\rangle (t).
\end{equation}
Following Ref. \cite{flindt1}, we write Eq. \eqref{mcd} by moving to Laplace space as 
\begin{equation}\label{mcd2}
    S(\omega)=-\omega^2 \mathfrak{Re}\left\{\langle\!\langle n^2 \rangle\!\rangle(z = -i\omega)\right\}.
\end{equation}
To find $\expval{\expval{ n^2 }}(z)$, we use the $\chi-$dependent density matrix in Laplace space 
\begin{equation}\label{rhoz}
    \rho(\chi,z) = \sum_{n=0}^{\infty} \{\Omega_z [(e^{i\chi}-1)J_{\text{R}} + (e^{-i\chi}-1)J_{\text{L}}]\}^n \Omega_z \rho_{\text{st}},
\end{equation}
where the propagator in Laplace space is defined by $\Omega_z=[z-L(\chi=0)]^{-1}$. The second cumulant is found by expanding Eq. \eqref{rhoz} in $\chi$ and keeping only the second-order terms
\begin{equation}\label{n2z}
\begin{split}
   \langle\!\langle n^2 \rangle\!\rangle (z)&= \partial_{i\chi}^2 \text{Tr}\{\rho(\chi,z) \}|_{\chi \to 0} \\[0.2em]&= \text{Tr} \{( \Omega_z \mathcal{J}\Omega_z +2 \Omega_z\mathcal{I} \Omega_z \mathcal{I} \Omega_z) \rho_{\text{st}} \},
\end{split}
\end{equation}
where $\mathcal{J}=J_{\text{R}}+J_{\text{L}}$ and $\mathcal{I}=J_{\text{R}}-J_{\text{L}}$, where $J_{\text{L}}$ and $J_{\text{R}}$ are defined below Eq. \eqref{mec}. 
\subsubsection{Low-drive limit}
We obtain an analytical expression for the FFN in the low-drive limit. The steady state is given from Eqs. \eqref{pertz}-\eqref{perts} by putting $\chi=0$, resulting in the FFN 
\begin{equation}\label{lowfp}
    S(\omega)=\Gamma_{\text{LR}}+\Gamma_{\text{RL}}+\frac{2\Gamma_{\text{RR}}\omega^2}{\Gamma_{g0}^2+\omega^2}.
\end{equation}
The frequency dependence is governed by the rate $\Gamma_{g0}$. This can be understood in the following way: In the low-drive limit, the system is almost always in its ground state - the event of an electron getting excited and jumping out of the DQD occurs with small probability. If it does, another electron will tunnel into the ground state with rate $\Gamma_{g0}$. Tunneling events thus occur in pairs. Events within one pair are correlated and separated by the time-scale $1/\Gamma_{g0}$, which can thus be 
interpreted as the dead-time of the photodetector. Tunneling events from different pairs are uncorrelated, as the pairs are separated by a large time-scale proportional to $1/f^2$ [cf. Eq. \eqref{l0perf1}]. 

We note that in the limit of ideal photodetection, the FFN reduces to $S(\omega)=\dot{N}$, which is equivalent to the noise of photons emitted from a coherent light source.
\subsubsection{Large-drive limit}
It was pointed out above that the Liouvillian reduces to a $5\times 5$ matrix in the regime where the drive is strong. Inserting the Liouvillian into the formula for the propagator $\Omega_z$ and taking the limit $f\to\infty$ in Eq. \eqref{n2z} for both the propagator and the steady state, we find for the FFN
\begin{equation}\label{ffnlf}
\begin{split}
    S(\omega)&=\frac{1}{\Gamma_{0e}+2\Gamma_{g0}}\Big[\Gamma_{0e}\Gamma_{\text{R,in}}+\Gamma_{g0}\Gamma_{\text{R,out}}\\
    &-\frac{2(2\Gamma_{\text{R,in}}+\Gamma_{\text{R,out}})(\Gamma_{0e}^2\Gamma_{\text{R,in}}+2\Gamma_{g0}^2\Gamma_{\text{R,out}})}{(\Gamma_{0e}+2\Gamma_{g0})^2+4\omega^2}\Big].
\end{split}
\end{equation}
The frequency dependence is thus governed by both the in- and outgoing rate from the ground and excited state of the DQD. In direct analogy to the low-drive case we interpret the time-scale $1/(\Gamma_{0e}+2\Gamma_{g0})$ as the dead-time of the photodetector. We note however that in the large drive limit, the detector is oversaturated and cannot detect every single photon.
\section{Mean-field approach}\label{mfp}
To obtain analytical expressions for the different cumulants of the photocurrent beyond the low- and large-drive limit, we employ a mean-field approach \cite{plank, prataviera}. Albeit being an approximation, we find that the mean-field solutions reproduce the exact results to a large degree and in a wide range of parameters. We note that this approach has been used in a similar system to obtain the average current through the DQD \cite{vavilov}. Here we extend that work by analyzing the zero-frequency FCS and the FFN.

Our approach relies on the assumption that correlations between the states of the DQD and the resonator can be neglected, such that at all times we can approximate the system’s density matrix as $\rho(t) = \rho_{\text{DQD}}(t) \otimes \rho_{\text{r}}(t)$, a product between the reduced density matrix for the DQD and the resonator. Inserting this factorization of the density matrix into Eq. \eqref{lme} and tracing over the degrees of freedom of the resonator, we get a master equation for the DQD, where the properties of the resonator enter only via their average values. We thereafter include the counting fields into this master equation, as we want to count electron-tunneling events through the DQD, and get 
\begin{align}
    \partial_t \rho_{\text{DQD}} &\! \begin{aligned}[t]&= -i[\bar{H}_{\text{DQD}},\rho_{\text{DQD}}]\\[0.2em]& + \Gamma_{\text{L,in}} \mathcal{D}[s_g^{\dagger}]\rho_{\text{DQD}}+ \Gamma_{\text{L,out}} \mathcal{D}[s_e]\rho_{\text{DQD}}\\[0.2em]
    &+\Gamma_{\text{R,in}}  (e^{-i\chi}s_g^{\dagger}\rho_{\text{DQD}} s_g -\tfrac{1}{2} \{s_gs_g^{\dagger},\rho_{\text{DQD}}\})\\[0.2em]&+   \Gamma_{\text{R,out}}(e^{i\chi}s_e\rho_{\text{DQD}} s_e^{\dagger}-\tfrac{1}{2}\{s_e^{\dagger}s_e,\rho_{\text{DQD}}\})\\[0.2em]
    &+ \frac{\gamma_{\phi}}{2}\mathcal{D}[\sigma_z]\rho_{\text{DQD}} +\gamma_- \mathcal{D}[\sigma_-]\rho_{\text{DQD}},
\end{aligned} \label{mecd}\\[0.8em]
    \bar{H}_{\text{DQD}}&\! \begin{aligned}[t]&= \Delta \frac{\sigma_z}{2} + g\left(\expval{a^{\dagger}}\sigma_-+\expval{a}\sigma_+\right).
\end{aligned} \label{hdqdmf}
\end{align}
Along the same lines, tracing over the degrees of freedom of the DQD gives a master equation for the resonator
\begin{align}
     \partial_t \rho_{\text{r}} &\! \begin{aligned}[t]&= -i[\bar{H}_{\text{r}},\rho_{\text{r}}] +\kappa \mathcal{D}[a]\rho_{\text{r}},
\end{aligned} \label{mecc} \\[0.8em]
    \bar{H}_{\text{r}} &\! \begin{aligned}[t]&= \Delta a^{\dagger}a +g\left(a^{\dagger}\expval{\sigma_-}_{\chi=0}+a\expval{\sigma_+}_{\chi=0}\right)\\
    & +f\left(a^{\dagger}+a\right). 
\end{aligned} \label{hrmf}
\end{align}
We use Eqs. \eqref{mecd}-\eqref{hrmf} to find equations of motions for operators which act on the subspace of the DQD and the resonator, respectively.
\subsection{Mean-field FCS}
In the long-time limit, we write the reduced, $\chi-$dependent density matrix for the DQD as
\begin{equation}
    \tilde{\rho}_{\text{DQD}}(\chi) \equiv \lim_{t\to\infty} \frac{\rho_{\text{DQD}}(\chi,t)}{\text{Tr}\{\rho_{\text{DQD}}(\chi, t)\}}. \label{reddqd}
\end{equation}
With this at hand, we obtain the following set of equations for the DQD [cf. Eq. \eqref{link}]
\begin{align}
    \lambda_0(\chi) \tilde{p}_0 &\! \begin{aligned}[t]&= -\Gamma_{g0} \tilde{p}_0 + (\Gamma_{\text{L,out}}+ e^{i\chi}\Gamma_{\text{R,out}}) \tilde{p}_e,  
\end{aligned} \label{p0}\\[0.8em]
    \lambda_0(\chi) \tilde{p}_e &\! \begin{aligned}[t]&= -(\Gamma_{0e}+\gamma_-) \tilde{p}_e \\[0.2em]&- ig(\expval{a}\expval{ \sigma_+}-\expval{a^{\dagger}}\expval{ \sigma_-}),  
\end{aligned} \label{pe}\\[0.8em]
    \lambda_0(\chi) \tilde{p}_g &\! \begin{aligned}[t]&= (\Gamma_{\text{L,in}}+e^{-i\chi}\Gamma_{\text{R,in}}) \tilde{p}_0 + \gamma_-\tilde{p}_e\\[0.2em]&+ ig(\expval{a}\expval{ \sigma_+}-\expval{a^{\dagger}}\expval{ \sigma_-}), 
\end{aligned} \label{pg}\\[0.8em]
    \lambda_0(\chi) \expval{\sigma_+} &\! \begin{aligned}[t]&= -\left(\frac{\tilde{\Gamma}}{2}  - i\Delta\right)  \expval{\sigma_+} - ig \expval{a^{\dagger}}\expval{ \sigma_z},
\end{aligned} \label{sp}
\end{align}
where $\tilde{p}_i=\text{Tr}\{\ketbra{i}{i}\tilde{\rho}_{\text{DQD}}(\chi)\}$ and $\expval{\sigma_j}=\text{Tr}\{\sigma_j\tilde{\rho}_{\text{DQD}}(\chi)\}$ for $j \in \{+,-,z\}$. For the resonator, we have
\begin{equation}\label{ad}
    0= -\left(\frac{\kappa}{2}-i\Delta\right)\expval{a^{\dagger}}+if+ig\expval{\sigma_+}_{\chi=0}.
\end{equation}

Using Eqs. \eqref{p0}-\eqref{ad}, together with the conservation of the effective probabilities $\tilde{p}_0+\tilde{p}_e+\tilde{p}_g=1$, we find two coupled equations for $\tilde{p}_e$ and $\lambda_0(\chi)$
\begin{widetext}
\begin{equation}\label{pe1}
     \tilde{p}_e=\frac{-16f^2g^2( \tilde{\Gamma}^2+4\Delta^2)(\tilde{\Gamma}+2\lambda_0)(\Gamma_{g0}+\lambda_0)\Lambda}{(\Gamma_{0e}+\gamma_-+\lambda_0)[(\tilde{\Gamma}+2\lambda_0)^2+4\Delta^2][(\Gamma_{g0}+\lambda_0)^2(\tilde{\Gamma}^2+4\Delta^2)(\kappa^2+4\Delta^2)+8g^2\Lambda(\Gamma_{g0}+\lambda_0)(4\Delta^2-\tilde{\Gamma}\kappa)+16g^4\Lambda^2]},
\end{equation}
\begin{equation}\label{l0}
\lambda_0=\frac{\tilde{p}_e}{\Gamma_{g0}+\lambda_0}[(e^{i\chi}-1)(\Gamma_{\text{L,in}}\Gamma_{\text{R,out}}+\Gamma_{\text{R,out}}\lambda_0)+(e^{-i\chi}-1)\Gamma_{\text{R,in}}\Gamma_{\text{L,out}}],
\end{equation}

where $\Lambda \equiv (2\tilde{p}_e-1)\Gamma_{g0}-\lambda_0+\tilde{p}_e(\Gamma_{\text{L,out}}+e^{i\chi}\Gamma_{\text{R,out}}+2\lambda_0)$.

The solution of Eq. \eqref{l0} which goes to zero for $\chi\to 0$ is the one of interest. It is given by

\begin{equation}\label{l01}
    \lambda_0(\chi)= \frac{1}{2} \Big\{ -\Gamma_{g0} + (e^{i\chi}-1)\tilde{p}_e\Gamma_{\text{R,out}}+
    \sqrt{4\tilde{p}_e[(e^{i\chi}-1)\Gamma_{\text{L,in}}\Gamma_{\text{R,out}}+(e^{-i\chi}-1)\Gamma_{\text{R,in}}\Gamma_{\text{L,out}}] +[\Gamma_{g0}-(e^{i\chi}-1)\tilde{p}_e\Gamma_{\text{R,out}}]^2}
     \Big\}.
\end{equation}
\end{widetext}
By expanding Eq. \eqref{l01} in $\chi$ using  $\lambda_0(\chi)\approx i\chi \langle\!\langle n^1\rangle\!\rangle-\frac{\chi^2}{2}\langle\!\langle n^2\rangle\!\rangle+\mathcal{O}[\chi^3]$ and $\tilde{p}_e \approx \tilde{p}_e^{(0)} + i \chi \tilde{p}_e^{(1)} + \mathcal{O}[\chi^2]$, the first and second cumulant, which give the mean current and the shot-noise, respectively, are

\begin{align}
    \langle\!\langle n^1\rangle\!\rangle&\! \begin{aligned}[t]&= \frac{\tilde{p}_e^{(0)}}{\Gamma_{g0}}(\Gamma_{\text{L,in}}\Gamma_{\text{R,out}}-\Gamma_{\text{R,in}}\Gamma_{\text{L,out}}),
\end{aligned} \label{n1}\\[0.8em]
  \langle\!\langle n^2\rangle\!\rangle  &\! \begin{aligned}[t]&=\frac{2\tilde{p}_e^{(1)}}{\Gamma_{g0}}(\Gamma_{\text{L,in}}\Gamma_{\text{R,out}}-\Gamma_{\text{R,in}}\Gamma_{\text{L,out}})\\[0.2em]&+\frac{2\tilde{p}_e^{(0)2}}{\Gamma_{g0}^3}[\Gamma_{0e}\Gamma_{\text{R,in}}(\Gamma_{\text{L,in}}\Gamma_{\text{R,out}}-\Gamma_{\text{R,in}}\Gamma_{\text{L,out}})]\\[0.2em]&+\frac{\tilde{p}_e^{(0)}}{\Gamma_{g0}}(\Gamma_{\text{L,in}}\Gamma_{\text{R,out}}+\Gamma_{\text{R,in}}\Gamma_{\text{L,out}}).
\end{aligned} \label{n2}
\end{align}
Third and higher-order cumulants could be found along the same lines. The different orders of $\tilde{p}_e$ are found by expanding Eq. \eqref{pe1} to zeroth and first order in $\chi$, respectively. The full expressions for $\tilde{p}_e^{(0)}$ and $\tilde{p}_e^{(1)}$ are given in Appendix \ref{ABB}.

In the low-drive regime we can again obtain a closed analytical expression for the FCS by expanding Eqs. \eqref{pe1}-\eqref{l0} to lowest order in $f$. This gives

\begin{equation}\label{l0mf}
    \lambda^{\text{mf}}_0(\chi)=\Gamma_{\text{LR}}^{\text{mf}}(e^{i\chi}-1)+\Gamma_{\text{RL}}^{\text{mf}}(e^{-i\chi}-1),    
\end{equation}
where 
\begin{equation}\label{gmf}
    \frac{\Gamma_{\alpha\beta}^{\text{mf}}-\Gamma_{\alpha\beta}}{\Gamma_{\alpha\beta}}=\frac{8g^2\kappa\gamma_{\phi}}{(\Gamma_{0e}+\gamma_-)(\tilde{\Gamma}+\kappa)(4g^2+\tilde{\Gamma}\kappa)},
\end{equation}
where $\alpha,\beta\in\{\text{L, R}\}$. We again get a bi-directional Poisson process describing the statistics of the photocurrent in the low-drive regime but with different rates compared to the perturbative result in Eq. \eqref{l0perf}.

The FCS in the large-drive regime is found by expanding Eqs. \eqref{pe1}-\eqref{l0} to zeroth order in $1/f$. This gives exactly the same result as in Eq. \eqref{lf}, i.e. the mean-field approach is exact in the large-drive limit.

In Fig. \ref{fig:prob}a) we plot the first two cumulants in Eqs. \eqref{n1}-\eqref{n2} and compare them to the low-drive result in Eq. \eqref{l0mf} and the large-drive result in Eq. \eqref{lf}. We can clearly see that the cumulants match the low-drive limit in the regime where the drive is weak and approach the large-drive limit asymptotically. 

In addition, we look at the non-Gaussian properties of the probability distribution $p(n,t)$ by plotting the third cumulant, see Fig. \ref{fig:prob}b). We see that the third cumulant is non-zero and is monotonically increasing as a function of the microwave drive. This shows a non-Gaussian behavior of the distribution. In the inset we plot the logarithm of the probability distribution in the saddle-point approximation (see Appendix \ref{AC} for details) for the low- and large-drive limit and compare them to the logarithm of a Gaussian distribution. For the chosen parameters, we see that both the low- and large-drive limits are skewed and hence deviate from a Gaussian distribution.
\subsection{Mean-field FFN}
The mean-field approach also allows us to find an analytical solution for the FFN. The Liouvillian for the DQD is a $5\times5$ matrix in the basis of the DQD. The full expression for the noise is, however, too lengthy to be given here. We find a closed analytical expression in the low-drive regime by expanding Eq. \eqref{n2z} to lowest order in $f$ 
\begin{equation}\label{ffnmf}
    S^{\text{mf}}(\omega)=\Gamma^{\text{mf}}_{\text{LR}}+\Gamma^{\text{mf}}_{\text{RL}}+\frac{2\Gamma^{\text{mf}}_{\text{RR}}\omega^2}{\Gamma_{g0}^2+\omega^2}.
\end{equation}
Similar to the mean-field zero-frequency FCS in Eq. \eqref{l0mf}, the form of the expression for the FFN is the same as in Eq. \eqref{lowfp}, however with the rates given by their mean-field expression in Eq. \eqref{gmf}. In the large-drive limit, the FFN coincides with the exact result in Eq. \eqref{ffnlf}. 
\begin{figure*}[ht!]
\centering
    \includegraphics[width=1\textwidth]{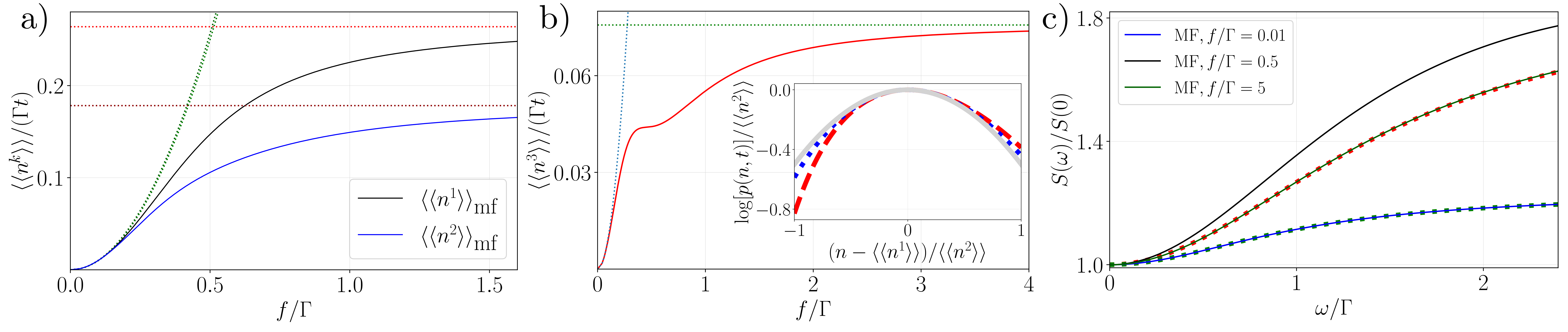}
    \caption{a) The first two zero-frequency cumulants within the mean-field approximation (full lines) normalized with $\Gamma t$ and plotted as a function of the normalized drive amplitude $f/\Gamma$, where $\Gamma\equiv\Gamma_{\text{L}}=\Gamma_{\text{R}}$. The cumulants are compared to the analytical results in the low- and large-drive limits (dotted lines). b) The third cumulant within the mean-field approximation (full line) normalized with $\Gamma t$ and plotted as a function of $f/\Gamma$. The third cumulant is also compared to the low- and large-drive limits (dotted lines). Inset: Probability distribution in the saddle-point approximation for the low-drive (red-dashed line) and the large-drive limit (blue dotted line) as a function of $(n-\langle\!\langle n^1\rangle\!\rangle)/\langle\!\langle n^2\rangle\!\rangle$. A Gaussian is shown for comparison (gray full line). The distributions are normalized by their second cumulant $\langle\!\langle n^2\rangle\!\rangle$. c) The FFN within the mean-field approximation (full lines) normalized with its value at zero frequency $S(0)$ and plotted as a function of the normalized frequency $\omega/\Gamma$ for different values of $f/\Gamma$. The FFN is compared to the analytical results in low- and large-drive limit (dotted lines). 
    The chosen system parameters are $g/\Gamma=0.457$, $\kappa_{\text{in}}/\Gamma=0.094$, $\kappa/\Gamma=0.337$, $\epsilon/\Gamma=-17.57$, $t_c/\Gamma=6.78$,  $\gamma_-/\Gamma=0.5$, $\gamma_{\phi}/\Gamma=3.92$, and $\Delta = 0$. The numerical values for the rates are taken from Ref. \cite{khan}.}
    \label{fig:prob}
\end{figure*}
In Fig. \ref{fig:prob}c) we plot the noise for different values of $f$ and compare them to the low-drive result in Eq. \eqref{ffnmf} and the large-drive result in Eq. \eqref{ffnlf}. We can see that the mean-field FFN matches the low-drive limit well for small values of $f$ and approaches the large-drive limit for larger values of $f$.
\subsection{Validity of the mean-field approach}
It is a priori not clear to what extent the assumption to neglect correlations between the DQD and the resonator is justified. We therefore check the validity of the mean-field approach by first comparing the result in Eq. \eqref{l0mf} to the perturbative one in in Eq. \eqref{l0perf}. We have seen that the mean-field results are exact in the large-drive limit. From Eq. \eqref{gmf} it is clear that the mean-field result in the low-drive limit becomes exact if either $g=0$, $\gamma_{\phi}=0$ or $\kappa=0$. For $g=0$, the DQD and the resonator are not coupled and thus no correlations can arise. For
$\kappa=0$ or $\gamma_{\phi}=0$, we find that the state is not of the product form assumed in the mean-field approach, but that $\expval{a^{\dagger}\sigma_-}$ factorizes in the low-drive limit. As can be anticipated from Eqs. \eqref{pe} and \eqref{pg}, this factorization is sufficient for the mean-field approach to become exact. For $\gamma_{\phi}=0$, we may understand this factorization because the two-level system provided by the DQD containing one electron can be treated as a harmonic oscillator in the low-drive limit (where only the two lowest energy levels are relevant). In the absence of dephasing, and if there is an electron in the DQD, the dot-resonator system may then be described as two bosonic modes (one of which is coherently driven), coupled by a beamsplitter interaction which is well known to preserve the product state structure for coherent states \cite{scott}. Correlations between the DQD and the resonator still arise because electrons are entering and leaving the DQD. The empty DQD state however does not affect $\expval{a^{\dagger}\sigma_-}$. For $\kappa=0$, we have not been able to find a compelling physical picture explaining the factorization of $\expval{a^{\dagger}\sigma_-}$.

We note that the photo-detector requires a non-zero $g$ and $\kappa$ to operate. Importantly, in the limit of ideal photodetection, where $\Gamma_{\text{L,R}} \gg \gamma_-, \gamma_{\phi}$, the difference in Eq. \eqref{gmf} becomes vanishingly small. 

Outside the low- and large-drive limit, the mean-field solutions can be benchmarked against an exact, numerical result for the cumulants. Details on the numerical calculations can be found in Appendix \ref{AD}. In Fig. \ref{fig:cumulants}, we compare the first two zero-frequency cumulants in Eqs. \eqref{n1} and \eqref{n2} and the FFN from the mean-field calculations with exact numerics. We can clearly see that, in the chosen parameter regime, the mean-field approach fits the numerics very well.

\begin{figure*}[ht]
\centering
    \includegraphics[width=0.899\textwidth]{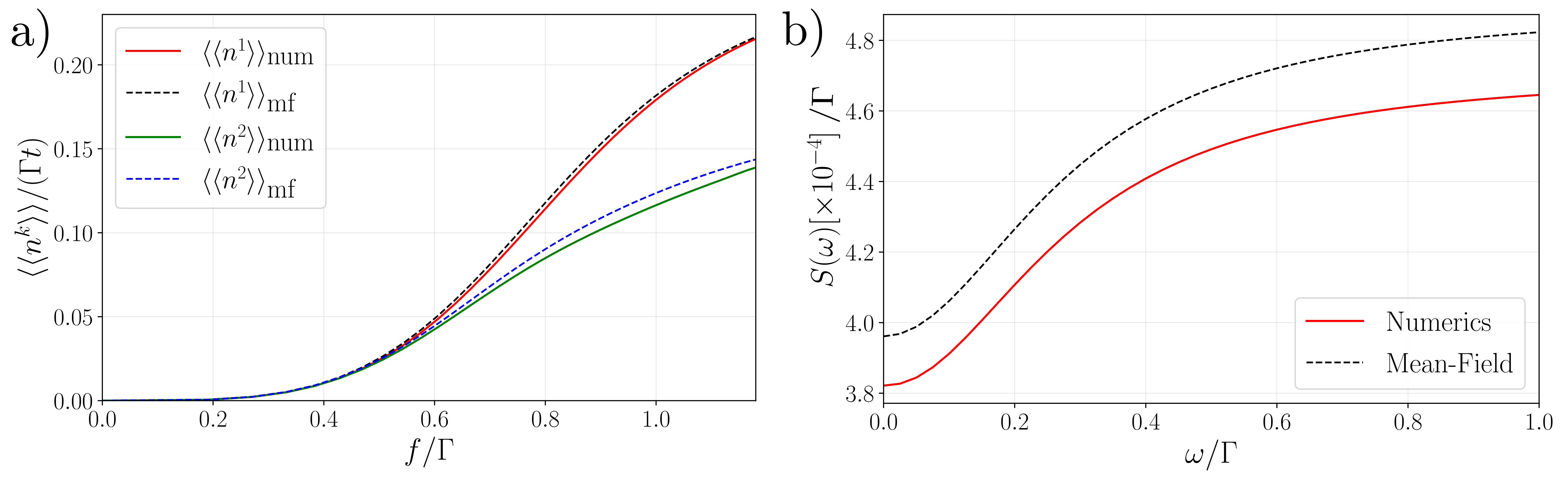}
    \caption{a) Comparisons between analytical (dashed lines) and numerical (full lines) calculations of the normalized, zero-frequency cumulants $\langle\!\langle n^k\rangle\!\rangle/(\Gamma t)$ for $k=1,2$, where $\Gamma \equiv \Gamma_{\text{L}}=\Gamma_{\text{R}}$. The cumulants are plotted as a function of the normalized drive amplitude $f/\Gamma$. b) Comparison between analytical (dotted line) and numerical (full line) calculations of the normalized FFN $S(\omega)/\Gamma$ plotted as a function of the normalized frequency $\omega/\Gamma$. The analytical solutions of the cumulants and the noise were found using a mean-field approach, and the numerical solutions were found using QuTip \cite{qutip}. The chosen system parameters are $\Gamma t=2.9\cdot10^5$, $g/\Gamma=0.457$, $\kappa_{\text{in}}/\Gamma=0.094$, $\kappa/\Gamma=0.337$, $\epsilon/\Gamma=-17.57$, $t_c/\Gamma=6.78$,  $\gamma_-/\Gamma=0.5$, $\gamma_{\phi}/\Gamma=3.92$, $\Delta = 0$ for a) and also $f/\Gamma=0.036$ for b). The numerical values for the rates are taken from Ref. \cite{khan}.}
    \label{fig:cumulants}
\end{figure*}

\section{Conclusions and outlook}\label{conc}

To summarize, we have theoretically investigated the FCS of the photocurrent through a DQD which is coupled to a driven microwave resonator. The zero-frequency FCS and the FFN show that the detector preserves the statistics of the incoming microwave photons in the parameter regime where the system was found to be an ideal photodetector. The statistics in that case can be described by a Poisson process. We have found analytical solutions for the zero-frequency FCS and the FFN in the limit of low and large applied drive, respectively. We have also shown how a mean-field approximation can be used to compute the zero-frequency cumulants and the FFN. Analysis of the third cumulant showed that the probability distribution $p(n,t)$ exhibits a non-Gaussian behavior. Comparison of the mean-field approach to exact numerics showed excellent agreement in the parameter regime of interest.

Our results pave the way for a better theoretical understanding of the statistical properties of electrons and photons inside a DQD-resonator hybrid structure used as an efficient and continuous single-photon detector. To achieve single-photon detection, real-time charge sensing techniques could be employed \cite{gustav, sha}. Such techniques typically induce back-action from the measurement device \cite{sha}, affecting the statistics of the photocurrent. A theoretical investigation into a specific architecture allowing for single-photon detection provides an interesting avenue for future research. Other interesting directions include the investigation of different drives, for instance single-photon sources, as well as the possibility of creating non-classical states of light and the study of their statistical properties using the present architecture \cite{walls,liu,maisi,eichler,tineke,segal}. 

\section*{Acknowledgments}

This work was supported by the Knut and Alice Wallenberg
Foundation through the Wallenberg Center for Quantum Technology (WACQT). P. P. P. acknowledges funding from the Swiss National Science Foundation (Eccellenza Professorial Fellowship PCEFP2\_194268).

\appendix
\section{Expression for the density matrix in the long-time limit}\label{AA}
In order to derive the expression for the $\chi-$dependent density matrix in the long-time limit and find the explicit expression for $\mu_0(\chi)$, we have to start by giving the eigendecomposition of the the Liouvillian $L(\chi)$. Since the Liouvillian is a superoperator, its eigenvectors will be operators. We represent these eigenvectors using a special Dirac notation and write $\ket{\psi_k(\chi)}\!\rangle$ for the right and $\langle\! \bra{\psi_k(\chi)}$ for the left eigenvector, respectively \cite{flindt2}. Note that, in general, $\ket{\psi_k(\chi)}\!\rangle^{\dagger}\neq\langle\! \bra{\psi_k(\chi)}$. Using this, the eigendecomposition of $L(\chi)$ is given by 
\begin{equation}\label{eigdec}
    L(\chi)=\sum_{k=0}^{N-1}\lambda_k(\chi)\ket{\psi_k(\chi)}\!\rangle\langle\! \bra{\psi_k(\chi)},
\end{equation}
where $\lambda_k(\chi)$ are the eigenvalues of $L(\chi)$. Using Eq. \eqref{eigdec}, we can write Eq. \eqref{sol} of the main text as
\begin{equation}\label{ed2}
    \rho(\chi,t)=\sum_{k=0}^{N-1} e^{\lambda_k(\chi)t}\ket{\psi_k(\chi)}\!\rangle\langle\! \bra{\psi_k(\chi)}\rho_{\text{st}}.
\end{equation}
We have seen that, in the long-time limit, only the eigenvalue $\lambda_0(\chi)$ will be of importance, since it yields the smallest exponentially damped contribution to the density matrix. Using Eq. \eqref{ed2} together with Eq. \eqref{sol} of the main text, we can write the $\chi -$dependent density matrix in the long-time limit as
\begin{equation}\label{ed3}
   \lim_{t\to\infty} \rho(\chi,t)=e^{\lambda_0(\chi)t}\mu_0(\chi),
\end{equation}
where 
\begin{equation}\label{mu0}
    \mu_0(\chi)= \ket{\psi_0(\chi)}\!\rangle\langle\! \bra{\psi_0(\chi)}\rho_{\text{st}}.
\end{equation}

\section{Perturbative low-drive expansion}\label{AB}
In this Appendix, we will give the explicit expressions for the normalized $\chi -$dependent density matrix in the long-time limit $\tilde{\rho}(\chi)$ to different orders in $f$. To zeroth order, there are no photons arriving in the resonator, such that the electron on the DQD will remain in the ground state for ever. We can thus make the following ansatz
\begin{equation}\label{ss0}
    \tilde{\rho}^{(0)}(\chi)=\ketbra{g,0_{\text{r}}}{g,0_{\text{r}}},
\end{equation}
where $\ket{0_{\text{r}}}$ describes the state of the resonator with zero photons. We have added the subscript in order to differentiate the empty-resonator state from the empty-dot state $\ket{0}$.

To first order, we consider superpositions of states having one and zero excitations, either in the resonator or in the DQD. We thus make the following ansatz
\begin{equation}\label{ss1}
\begin{split}
\tilde{\rho}^{(1)}(\chi)&=\alpha_1 \ketbra{g,1}{g,0_{\text{r}}}+\alpha_2 \ketbra{g,0_{\text{r}}}{g,1}\\& + \alpha_3 \ketbra{e,0_{\text{r}}}{g,0_{\text{r}}} + \alpha_4 \ketbra{g,0_{\text{r}}}{e,0_{\text{r}}} .
\end{split}
\end{equation}
The prefactors $\alpha_{1-4}$ can be found by plugging Eqs. \eqref{ss1} and \eqref{ss0} into Eq. \eqref{pertf} of the main text.

To second order, we similarly include all density matrix elements where the number of excitations in the bra and the ket sum to two, resulting in the following ansatz
\begin{equation}\label{ss2}
\begin{split}
\hspace*{-0.35cm}
    \tilde{\rho}^{(2)}(\chi)&=\beta_1 \ketbra{g,1}{g,1} + \beta_2 \ketbra{e,0_{\text{r}}}{e,0_{\text{r}}} + \beta_3 \ketbra{g,1}{e,0_{\text{r}}} \\&+ \beta_4 \ketbra{e,0_{\text{r}}}{g,1}+ \beta_5 \ketbra{g,2}{g,0_{\text{r}}}+\beta_6 \ketbra{g,0_{\text{r}}}{g,2} \\&+\beta_7 \ketbra{e,1}{g,0_{\text{r}}} + \beta_8 \ketbra{g,0_{\text{r}}}{e,1} + \beta_9 \ketbra{g,0_{\text{r}}}{g,0_{\text{r}}} \\&+ \beta_{10} \ketbra{0,0_{\text{r}}}{0,0_{\text{r}}}.
\end{split} 
\end{equation}
Note that we included the last two terms since they are coupled to the first and second term by the process of a photon or electron leaving the system, respectively. 
Plugging Eq. \eqref{ss2} into Eq. \eqref{perts} of the main text and using the fact that $\text{Tr}\{\tilde{\rho}^{(0)}(\chi)\} =1$ and $\text{Tr}\{\tilde{\rho}^{(1)}(\chi)\}=\text{Tr}\{\tilde{\rho}^{(2)}(\chi)\}=0$ gives us a closed set of equations for the prefactors $\beta_{1-10}$.
\section{Large-drive limit}\label{AAA}
We start by giving the equation of motion for the photonic annihilation operator in steady state [cf. Eq. \eqref{ad}]
\begin{equation}\label{a}
    0 = -\left(\frac{\kappa}{2}+i\Delta\right)\expval{a}-if-ig\expval{\sigma_-}.
\end{equation}
For large drives, fluctuations of the resonator photon state and the backaction from the DQD can be neglected. We thus make the replacement 
\begin{equation}\label{b}
    a=-\frac{2if}{\kappa+2i\Delta}.
\end{equation}
Inserting this replacement in to Eq. \eqref{mec} results in the following, $\chi-$dependent master equation
\begin{equation}\label{mec1}
\begin{split}
     \partial_t \rho &= -i[\tilde{H}',\rho] + \Gamma_{\text{L,in}} \mathcal{D}[s_g^{\dagger}]\rho+ \Gamma_{\text{L,out}} \mathcal{D}[s_e]\rho\\
     &+\Gamma_{\text{R,in}} (e^{-i\chi}s_g^{\dagger}\rho s_g - \tfrac{1}{2}\{s_gs_g^{\dagger},\rho\})\\&+  \Gamma_{\text{R,out}}(e^{i\chi}s_e\rho s_e^{\dagger}-\tfrac{1}{2}\{s_e^{\dagger}s_e,\rho\})\\
     &+ \frac{\gamma_{\phi}}{2}\mathcal{D}[\sigma_z]\rho +\gamma_- \mathcal{D}[\sigma_-]\rho,
\end{split}
\end{equation}
where
\begin{equation}\label{hprime}
    \tilde{H}' =   \Delta\frac{\sigma_z}{2} + 2ifg\left(\frac{\sigma_-}{\kappa-2i\Delta} -\frac{\sigma_+}{\kappa+2i\Delta} \right).
\end{equation}
\section{Analytical expressions for $\tilde{p}_e^{(0)}$ and $\tilde{p}_e^{(1)}$.}\label{ABB}
Expanding Eq. \eqref{pe1} to zeroth order in $\chi$ gives the following cubic equation for $\tilde{p}_e^{(0)}$
\begin{equation}\label{cubic}
    A\tilde{p}_e^{(0)3}+B\tilde{p}_e^{(0)2}+C\tilde{p}_e^{(0)}+D=0,
\end{equation}
where
\begin{widetext}
\begin{equation}\label{A}
    A=-16g^4(\Gamma_{0e}+\gamma_-)(\Gamma_{0e}+2\Gamma_{g0})^2,
\end{equation}
\begin{equation}\label{B}
    B=8g^2\Gamma_{g0}(\Gamma_{0e}+\gamma_-)(\Gamma_{0e}+2\Gamma_{g0})(\tilde{\Gamma}\kappa+4g^2-4\Delta^2),
\end{equation}
\begin{equation}\label{C}
    C=-\Gamma_{g0}\{\Gamma_{g0}(\Gamma_{0e}+\gamma_-)[(4g^2+\kappa\tilde{\Gamma}-4\Delta^2)^2+4\Delta^2(\kappa+\tilde{\Gamma})^2]+16g^2f^2\tilde{\Gamma}(\Gamma_{0e}+2\Gamma_{g0})\},
\end{equation}
\begin{equation}
    D=16f^2g^2\tilde{\Gamma}\Gamma_{g0}^2.
\end{equation}
A general solution to Eq. \eqref{cubic} is given by 
\begin{equation}\label{cubsol}
    p_e^{(0)}=\{q + [q^2 + (r-p^2)^3]^{\frac{1}{2}}\}^{\frac{1}{3}} + \{q - [q^2 + (r-p^2)^3]^{\frac{1}{2}}\}^{\frac{1}{3}} + p,
\end{equation}
where
\begin{align}
    p &=-\frac{B}{3A}, \label{p} \\
    q &= p^3 + \frac{BC-3AD}{6A^2}, \label{q} \\
    r &= \frac{C}{3A}. \label{r}
\end{align}
Two out of the three solutions in Eq. \eqref{cubsol} are complex and thus non-physical. We keep the real solution to compute the cumulants in Eqs. \eqref{n1}-\eqref{n2}. 

Expanding Eq. \eqref{pe1} to first order in $\chi$ gives the following linear equation for $\tilde{p}_e^{(1)}$
\begin{equation}\label{linsol}
\tilde{p}_e^{(1)}=-\frac{G\tilde{p}_e^{(0)}\langle\!\langle n^1\rangle\!\rangle(\tilde{\Gamma}^2+4\Delta^2)+E}{F(\tilde{\Gamma}^2+4\Delta^2)},
\end{equation}
where $E$, $F$, and $G$ are polynomials
\begin{equation}
E=16f^2g^2(A_1\tilde{p}_e^{(0)3}+B_1\tilde{p}_e^{(0)2}+C_1\tilde{p}_e^{(0)}+D_1),
\end{equation}
with coefficients 
\begin{equation}
    A_1=16g^4(\Gamma_{0e}+2\Gamma_{g0})^2\{\tilde{\Gamma}^3(\langle\!\langle n^1\rangle\!\rangle\Gamma_{0e}-\Gamma_{\text{R,out}}\Gamma_{g0})+4\Delta^2\{\langle\!\langle n^1\rangle\!\rangle[2\Gamma_{g0}(\Gamma_{0e}+2\Gamma_{g0})+\Gamma_{0e}\tilde{\Gamma}]-\Gamma_{\text{R,out}}\Gamma_{g0}\tilde{\Gamma}\}\},
\end{equation}
\begin{equation}
\begin{split}
    B_1&=32g^2\Gamma_{g0}^2(\Gamma_{0e}+2\Gamma_{g0})\{g^2\Gamma_{\text{R,out}}\Gamma_{g0}\tilde{\Gamma}(\tilde{\Gamma}^2+4\Delta^2)+\langle\!\langle n^1\rangle\!\rangle\{2\Gamma_{g0}\Delta^2(\Gamma_{0e}+2\Gamma_{g0})(4\Delta^2-\tilde{\Gamma}\kappa)
    \\&-g^2\{\Gamma_{0e}\tilde{\Gamma}^3+4\Delta^2[3\Gamma_{g0}(\Gamma_{0e}+2\Gamma_{g0})+\Gamma_{0e}\tilde{\Gamma}]\}\}\},
\end{split}
\end{equation}
\begin{equation}
\begin{split}
    C_1&=\Gamma_{g0}^2\{\Gamma_{\text{R,out}}\Gamma_{g0}\tilde{\Gamma}(\tilde{\Gamma}^2+4\Delta^2)[(\tilde{\Gamma}^2+4\Delta^2)(\kappa^2+4\Delta^2)-16g^4]+\langle\!\langle n^1\rangle\!\rangle\{16g^4\{\Gamma_{0e}\tilde{\Gamma}^3+4\Delta^2\\ &\times [6\Gamma_{g0}(\Gamma_{0e}+2\Gamma_{g0})+\Gamma_{0e}\tilde{\Gamma}]\}-128g^2\Delta^2(\Gamma_{0e}+2\Gamma_{g0})(4\Delta^2-\tilde{\Gamma}\kappa)-(\tilde{\Gamma}^2+4\Delta^2)(\kappa^2+4\Delta^2)\\ &\times \{\Gamma_{0e}\tilde{\Gamma}^3+4\Delta^2[\Gamma_{0e}\tilde{\Gamma}-2\Gamma_{g0}(\Gamma_{0e}+2\Gamma_{g0})]\}\}\},
\end{split}    
\end{equation}
\begin{equation}
    D_1=-8\langle\!\langle n^1\rangle\!\rangle\Gamma_{g0}^4\Delta^2[16g^4-8g^2(4\Delta^2-\tilde{\Gamma}\kappa)+(\tilde{\Gamma}^2+4\Delta^2)(\kappa^2+4\Delta^2)],
\end{equation}
and
\begin{equation}
    F=A_2\tilde{p}_e^{(0)4}+B_2\tilde{p}_e^{(0)3}+C_2\tilde{p}_e^{(0)2}+D_2\tilde{p}_e^{(0)}+E_2,
\end{equation}
with
\begin{equation}
    A_2=-16g^4(\Gamma_{0e}+2\Gamma_{g0})^2A,
\end{equation}
\begin{equation}
    B_2=-32g^4(\Gamma_{0e}+2\Gamma_{g0})^2B,
\end{equation}
\begin{equation}
\begin{split}
    C_2&=32g^4\Gamma_{g0}(\Gamma_{0e}+2\Gamma_{g0})^2\{\Gamma_{g0}(\Gamma_{0e}+\gamma_-)[48g^4+24g^2(\tilde{\Gamma}\kappa-4\Delta^2)+\tilde{\Gamma}^2(3\kappa^2+4\Delta^2)+4\Delta^2(\kappa^2+12\Delta^2)
    \\&-16\tilde{\Gamma}\kappa\Delta^2]-8f^2g^2\tilde{\Gamma}(\Gamma_{0e}+2\Gamma_{g0})\},
\end{split}
\end{equation}
\begin{equation}
\begin{split}
    D_2&=-16g^2\Gamma_{g0}^2(\Gamma_{0e}+2\Gamma_{g0})^2\{\Gamma_{g0}(\Gamma_{0e}+\gamma_-)\{64g^4+48g^4(\tilde{\Gamma}\kappa-4\Delta^2)+(\tilde{\Gamma}^2+4\Delta^2)(\tilde{\Gamma}\kappa-4\Delta^2)(\kappa^2+4\Delta^2)
    \\&-4g^2[16\tilde{\Gamma}\kappa\Delta^2-4\Delta^2(\kappa^2+12\Delta^2)-\tilde{\Gamma}^2(3\kappa^2+4\Delta^2)]\}-32f^2g^4\tilde{\Gamma}(\Gamma_{0e}+2\Gamma_{g0})\},
\end{split}
\end{equation}
\begin{equation}
\begin{split}
    E_2&=16f^2g^2\tilde{\Gamma}(\Gamma_{0e}+2\Gamma_{g0})[(\tilde{\Gamma}^2+4\Delta^2)(\kappa^2+4\Delta^2)-16g^4]+\Gamma_{g0}(\Gamma_{0e}+\gamma_-)\{16g^2\{16g^6-16g^4(4\Delta^2-\tilde{\Gamma}\kappa)
    \\&+(\tilde{\Gamma}^2+4\Delta^2)(\kappa^2+4\Delta^2)(\tilde{\Gamma}\kappa-4\Delta^2)+2g^2[\tilde{\Gamma}^2(3\kappa^2+4\Delta^2)+4\Delta^2(\kappa^2+12\Delta^2)-16\tilde{\Gamma}\kappa\Delta^2]\}
    \\&+[(\tilde{\Gamma}^2+4\Delta^2)(\kappa^2+4\Delta^2)]^2\}
\end{split}    
\end{equation}
and
\begin{equation}
    G=A_3\tilde{p}_e^{(0)4}+B_3\tilde{p}_e^{(0)3}+C_3\tilde{p}_e^{(0)2}+D_3\tilde{p}_e^{(0)}+E_3,
\end{equation}
with
\begin{equation}
    A_3=(2g)^8(\Gamma_{0e}+2\Gamma_{g0})^4,
\end{equation}
\begin{equation}
    B_3=-2^8g^6\Gamma_{g0}(\Gamma_{0e}+2\Gamma_{g0})^3(\tilde{\Gamma}\kappa+4g^2-4\Delta^2),
\end{equation}
\begin{equation}
    C_3=32g^4\Gamma_{g0}^2(\Gamma_{0e}+2\Gamma_{g0})^2[48g^4+24g^2(\tilde{\Gamma}\kappa-4\Delta^2)+2(\tilde{\Gamma}\kappa-4\Delta^2)^2+(\tilde{\Gamma}^2+4\Delta^2)(\kappa^2+4\Delta^2)],
\end{equation}
\begin{equation}
    D_3=-16g^2\Gamma_{g0}^3(\Gamma_{0e}+2\Gamma_{g0})(\tilde{\Gamma}\kappa+4g^2-4\Delta^2)[(4g^2+\kappa\tilde{\Gamma}-4\Delta^2)^2+4\Delta^2(\kappa+\tilde{\Gamma})^2],
\end{equation}
\begin{equation}
    E_3=\Gamma_{g0}^4[(4g^2+\kappa\tilde{\Gamma}-4\Delta^2)^2+4\Delta^2(\kappa+\tilde{\Gamma})^2]^2.
\end{equation}
\end{widetext}
\section{Saddle-point approximation}\label{AC}
The saddle-point approximation is a widely used method providing an approximation to a probability distribution using the CGF \cite{butler}. It states that, given the CGF $C(\chi,t)$, the probability distribution can be approximated by [cf. Eq. \eqref{prob}]
\begin{equation}\label{spa}
    p(n,t)\approx \frac{1}{\sqrt{2\pi C''(\chi,t)|_{\chi=\chi^*}}}\exp[C(\chi^*,t)-in\chi^*],
\end{equation}
where $\chi^*$ is the solution to the saddle-point equation
\begin{equation}\label{spe}
    C'(\chi^*,t)-n=0,
\end{equation}
where the derivatives of the CGF are defined with respect to $i\chi$.
In Fig. \ref{fig:prob}b), we plot the logarithm of the different distributions. Within the saddle-point approximation, and to exponential accuracy, this logarithm can be approximated as 
\begin{equation}\label{log}
    \log[p(n,t)]\approx C(\chi^*,t)-in\chi^*.
\end{equation}
\section{Numerics}\label{AD}
For the numerical calculations of the different cumulants, we follow Refs. \cite{flindt2} and \cite{marcos}, where the authors give a method for the evaluation of the zero-frequency cumulants and FFN, respectively. These methods rely solely on matrix multiplications and are, therefore, ideal for numerical implementation. We briefly review the methods for completeness. We write the steady state using the special Dirac notation defined in Appendix \ref{AA} as $\rho_{\text{st}}=\ket{\psi_0}\!\rangle$ and the corresponding left eigenvector as $ \langle\! \bra{\psi_0}$. The inner product between these objects is given by $\langle\! \braket{\psi_0}{\psi_0}\!\rangle=\text{Tr}\{\rho_{\text{st}}\}=1$. Next, we give the projector onto the steady state $\mathcal{P}=\mathcal{P}^2=\ket{\psi_0}\!\rangle\langle\!\bra{\psi_0}$ as well as its complement $\mathcal{Q}=1-\mathcal{P}$. Note that the following relations hold $L\mathcal{P}=\mathcal{P}L=0$ and therefore $L=\mathcal{Q}L\mathcal{Q}$. With this at hand, we can define the pseudo-inverse of the Liouvillian $\mathcal{R}=\mathcal{Q}L^{-1}\mathcal{Q}$ in the subspace where $L$ is regular. Having defined the necessary quantities, we find the first two zero-frequency cumulants 
\begin{align}
    \langle\!\langle n^1\rangle\!\rangle&= \langle\!\!\mel{\psi_0}{\mathcal{I}}{\psi_0}\!\rangle,\label{n1num} \\
    \langle\!\langle n^2\rangle\!\rangle&= \langle\!\!\mel{\psi_0}{\mathcal{J}}{\psi_0}\!\rangle-2\langle\!\!\mel{\psi_0}{\mathcal{I}\mathcal{R}\mathcal{I}}{\psi_0}\!\rangle, \label{n2num}
\end{align}
and the FFN is given by
\begin{equation}\label{ffnnum}
    S(\omega)=\langle\!\!\mel{\psi_0}{\mathcal{J}}{\psi_0}\!\rangle-2\mathfrak{Re}\left\{\langle\!\!\mel{\psi_0}{\mathcal{I}\mathcal{R}(\omega)\mathcal{I}}{\psi_0}\!\rangle\right\},
\end{equation}
where $\mathcal{R}(\omega)=\mathcal{Q}(L+i\omega)^{-1}\mathcal{Q}$ is the frequency-dependent pseudo-inverse. We have computed the different cumulants using QuTip \cite{qutip}.

\bibliography{main.bib}

\end{document}